# Temperature dependence investigation of dissipation processes in strongly anisotropic high-temperature superconductors of Bi-Pb-Sr-Ca-Cu-O system synthesized using solar energy.

J.G.Chigvinadze[a], J.V.Acrivos[b], S.M.Ashimov[a], D.D.Gulamova[c], T.V.Machaidze[a], D.Uskenbaev[c].

a – E. Andronikashvili Institute of Physics, 0177 Tbilisi, Georgia, Tamarashvili st. 6, Tel.: 995 32 397924, Fax: 995 32 391494, jaba@iphac.ge
b – San Jose State University, San Jose CA 95192-0101, USA, Tel.: 408 924 4972, Fax: 408 924 4045, jacrivos@athens.sjsu.edu,
c – The Institute of Materials Science SPA "Physics-Sun" of Academy of Science, Republic of Uzbekistan, 700084, Tashkent, Uzbekistan, Tel.: 998 71-135-74-96, Fax: 998 71-135-42-91.

## Abstract

The investigation of temperature dependence of damping and period of vibrations of HTSC superconductive cylinder of Bi-Pb-Sr-Ca-Cu-O system suspended by a thin elastic thread and performing axial-torsional vibrations in a magnetic field at temperatures above the critical one for the main phase $T_c$=107 K were carried out. It was observed some "chaos" temperature region in the temperature interval 107-138 K, where it is seen separate ripples of dissipation and oscillation frequency. It is assumed that the "chaos" region could point to a possibility of existence of other magnetic and more high-temperature phases as single islands in a normal materials matrix.

## 1. Introduction

The Bi-Pb-Sr-Ca-Cu-O system is one of perspective materials from the point of view of high-temperature superconductivity (HTSC) [1] applications. It is characterized by the high critical temperature of superconductive transition $T_c$=107 K and high upper magnetic critical field $H_{c2}$ of the order of 150 T [2]. The contemporary technology of fabrication of HTSC makes it possible to change their critical parameters among them the critical current density $J_c$ [3], high value of which is also necessary for HTSC applications in one of the most perspective directions in the contemporary technique: such as strong-current energetic [4]**,** in systems for creating of controlled thermonuclear synthesis (in Tokomaks), and also in medical diagnostics.

The Bi-Pb-Sr-Ca-Cu-O system is characterized by such high critical temperature of superconductive transition $T_c$ that it remains superconductive at temperatures when thermal fluctuations play considerable role and their energy becomes comparable with the elastic energy of vortices and the pinning energy [5]. This creates prerequisites for phase transitions. Due to the layered critical structure and anisotropy, which are characteristic of HTSC, the conditions arise for appearance of different phases in the *B-T* (*B* is magnetic induction, *T* is temperature) diagram for them [6-17]. As example, in the $Bi_{1,7}Pb_{0,3}Sr_2Ca_2Cu_3O_{10-\delta}$ (2223), during the increase of outer magnetic field (at *T*=const), the 3-dimentional Abrikosov`s vortices undergo phase transition in the 2-dimentional *2D* vortices, so-called "pancake" ones. During this process the absorption of low-frequency vibrations – the logarithmic decrement of damping of a superconducting cylinder, suspended by a thin elastic thread and performing of axial-torsional vibrations, is stepwisely changed and fall down approximately on the two orders of value [14]. The reason for such sharp decrease of low-frequency vibrations is the stepwise increase of pinning force predicted by American theoreticians [16] and observed experimentally in work [17]. Such phase transitions, the *3D-2D* transitions, in the vortex matter of HTSC is stipulated by their layered crystal structure and strong anisotropy (the factor of anisotropy for this superconductive system is of the order of 3000 [14]). The other example of a phase transition in the vortex matter of HTSC is the melting of Abrikosov vortex lattice near $T_c$. About the critical temperature $T_c$ the Abrikosov vortex lattice begins the melting and along with it the dynamics of vortex continuum flow is essentially changed. In particular, it is sharply changed relaxation phenomena. At temperatures much lower then $T_c$ in HTSC it is observed long relaxation processes – a slow logarithmic decrease of captured magnetic flux [18-20]. The logarithmic character of relaxation is explained by the thermoactivated Anderson creep [21]. In the range of Abrikosov vortex melting – near $T_c$, the logarithmic character of relaxation is changed on the power one with exponent 2/3 [22].

Consequently, the investigations of phase transitions in vortex matter of HTSC are very important for understanding of processes taking place in these materials.

It should be stressed also that for the understanding of observed in HTSC processes, the decisive factor could be the study of processes taking place in superconducting samples in their normal state, i.e. at temperatures above $T_c$.

No less essential is the search of new HTSC phases with higher temperatures of transition $T_c$ in a superconductive state.

Due to its high sensitivity and resolution ability, the mechanical method of investigations of dissipation processes in type II superconductors is very useful its application for study of magnetic flux structure, search and observation of superconductive phase – the Abrikosov-Shubnikov phase and phase transitions in vortex matter of HTSC [23].

To study namely these problems is devoted this work.

## 2. Samples under investigation and experiment

As the appearance of superconductive phase structure and flow of magnetic flux and phase transitions are usually related with temperature, we should have possibility to measure the logarithmic decrement of damping and frequency of vibrations of superconductive cylinder suspended by a thin elastic thread and performing axial-torsional vibrations in an outer magnetic field at different temperatures, both below and above the superconductive transition temperature $T_c$. It is particularly important to carry out these investigations above $T_c$ to find out new HTSC phases.

The mechanical method of investigation of pinning and dissipation processes in superconductors gives this possibility because at transition in a superconductive state the Abrikosov vortices would appear in HTSC [24] and these pinned vortices stipulate the change of vibration frequency of a suspension system, and vortices teared off from pinning centers, cause the change of dissipation of vibrations of a superconducting cylinder. This way, the investigations of temperature dependence of frequency and dissipation of a suspension system could give possibility to study the above noted problems and among them make it possible to find out new magnetic and superconductive phases with higher critical temperatures $T_c$ of transition in the superconductive states if they would present in the HTSC samples under investigations.

Using this method, in this work it was studied the creation of Abrikosov vortex lattice structure, the kinetics of formation and disintegration of this structure and the Abrikosov vortex motion dynamics [25,26].

The temperature dependence of dissipation and frequency of the suspension system with a superconductive sample was measured by the following procedure. In the very beginning, we switched a sample into the superconductive state in a magnetic field or in its absence and cooled it down to the temperature of boiling liquid nitrogen, i.e. $T$=77 K. It was further measured the temperature dependence of dissipation and frequency of the suspension system with a cylindrical HTSC sample.

The samples used in our investigations were synthesized with application of solar energy [3,27] and superfast quenching melt technology [28-30]. This way it was manufactured amorphous precursors and their use gives possibility to manufacture of high-density low-porous texturized ceramics with given dimensions of grains [31,32]. Starting mixtures of nominal composition $Bi_{1.7}Pb_{0.3}Sr_2Ca_2Cu_3O_{10-\delta}$ were prepared from previously annealed at 600°C $Bi_2O_3$, $PbO$, $SrCO_3$, $CaO$, $CuO$ powders with purity nor less than "PFA"(pure for analysis). The synthesis in a melt and the following quenching was realized by using as a heating source the concentrated solar (beam) flux (CSF) [33] in solar furnaces of 3 kW power and in imitators of solar flux of URAN type. This provides the purity of aimed material due to the lack of impurities from melting installations and crucibles, very small heating and cooling inertia, and a high velocity of reaching the necessary temperature what decreases the evaporation of starting components. The concentrated solar flux generates ozone in the medium, surrounding the melt, what makes it possible to obtain the superstoichiometric oxygen concentration. Such medium should influence the increase of Cu (II) content and, correspondingly, the critical parameters of aimed material [34]. The melting of material is realized on a water-cooled aluminium mold. The amorphous state of precursors was reached by the quenching of a melt using the powderization method [33].

A phase composition was monitored by X-ray diffraction method with a help of DRON-UM1 diffractometer using $CuK_\alpha$ radiation and the diffractometer model Rigaku Co, Ltd., Tokyo, Japan. The critical temperature of superconductive transition $T_c$ was defined by measurements of electric resistance temperature dependence with four-contact method and the magnetic susceptibility temperature dependence using the mutual induction method [31] and by the mechanical method to study the temperature dependence of frequency and dissipation of a superconductive cylinder suspended by a thin thread and performing axial-torsional oscillation in a magnetic field.

Precursors were fabricated as pieces with dimensions up to 1 cm, plates with thicknesses no more than 0,3 mm and viskers with lengths up to 10 mm, Ø< 0,4 mm. The phase compositions of pieces and plates were presented by amorphous and crystal phases. The exact interpretation of crystal phase composition was impossible due to the lack of most part and unclear reflex manifestation. The needles (viskers) were practically amorphous.

With aim to define the influence of the starting state of precursor (a plate and viskers) on the phase formation at crystallization it was for comparison fabricated HTSC samples by both melt and super fast quenching technologies from one side, and the standard solid state phase reaction from other side. This way the fabricated nominal composition $Bi_{1,7}Pb_{0,3}Sr_2Ca_2Cu_3O_{10-\delta}$ samples synthesized by both melt and standard solid state phase reaction technologies were subjected to thermotreatment in similar conditions at $T_{annealing} = $ 850 C during 60 hours. The phase composition in the bulk of samples both on the base of amorphous viskers and on the base of glass-crystal plates is presented by mainly 2223 and 2212 phases. At the same time, on surfaces of visker samples it was established up to 90-93% of 2223 phase and in samples on the plate base – the main phase was 2212 and there were of 2223 phase traces.

The critical temperature of transition in the superconducting state of the main $Bi_{1,7}Pb_{0,3}Sr_2Ca_2Cu_3O_{10-\delta}$ phase measured by the mechanical method for equal to $T_c$=107 K.

The $T_c$=107 K it was showed also standard methods, like $R=f(T)$ and $\chi=f(T)$ measurements [31].

But in these samples as it was pointed above, it is apparently present other phases (for examples it is as a rule certainly present the 2212 phase).

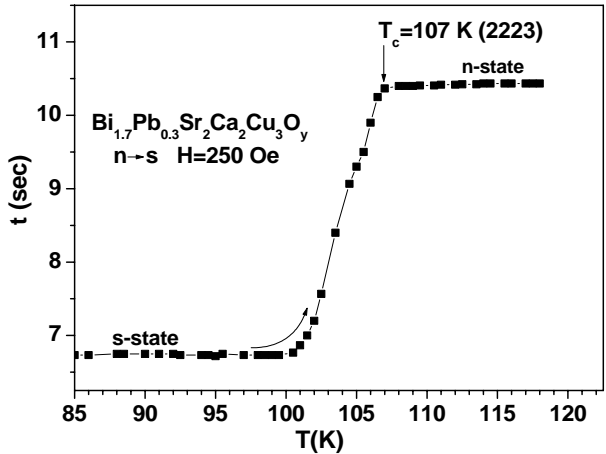

Fig.1. The oscillation period dependence of a superconducting $Bi_{1.7}Pb_{0.3}Sr_2Ca_2Cu_3O_{10-\delta}$ cylinder in the magnetic field H=250 Oe on temperature $T$.

In Fig.1, it is presented the dependence of oscillation period of a superconductive $Bi_{1.7}Pb_{0.3}Sr_2Ca_2Cu_3O_{10-\delta}$ system cylinder suspended by a thin elastic thread and performing the axial-torsional oscillation in a magnetic field directed perpendicular to the axis of cylinder on temperature with the reduction of temperature at the transition of a sample in the superconducting state the Abrikosov vortices are formed inside it which due to the interaction with an outer magnetic field create a mechanical momentum which in its turn increases the oscillation frequency of suspense system and consequently reduces the oscillation period. The period dependence curve $t=f(T)$ makes it possible to define the critical transition temperature of a sample from the normal to the superconductive state, i.e. we have one more method to define $T_c$. The critical temperature of superconducting transition defined this way coincides with value $T_c$=107 K defined by other methods ( by $R$ and $\chi$ ) [31].

### 3. Results and discussions

Due to the fact that the aim of this work was the investigation of dissipation processes in strongly anisotropic HTSC in the range of superconducting transition temperatures and above it, we have beforehand investigated this problem on usual (not strongly anisotropic) HTSC of $ErBa_2Cu_3O_{7-\delta}$ system, with the critical temperature $T_c$=92 K.

In Fig.2 it is presented the temperature dependence of period $t$ and logarithmic decrement of damping $\delta$ on temperature in the temperature range from helium 4,2 K to 110 K, and static magnetic field $H$=150 oe.

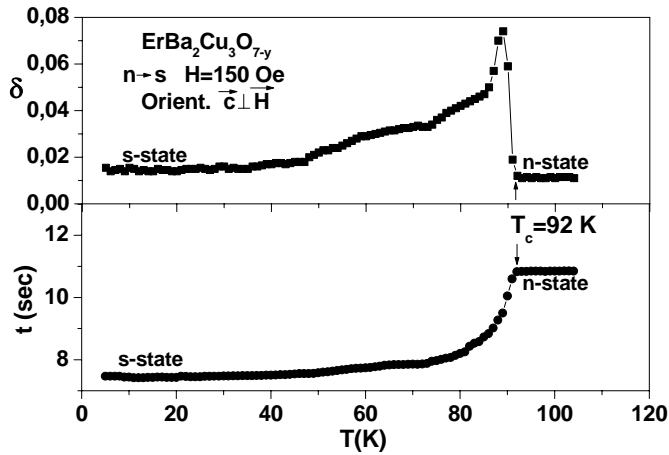

Fig.2. The dependence of period and logarithmic decrement of damping of oscillation on temperature of high-temperature superconductor of $ErBa_2Cu_3O_{7-\delta}$ system in a static magnetic field H=150 Oe.

The presented dependences in fig.2 are characteristic for HTSC of (123) system. The $\delta=f(T)$ dependence at constant magnetic field $H>H_{c1}$ for $ErBa_2Cu_3O_{7-\delta}$, as well as also for other HTSC of 123 system, reveals a typical maximum in $T_c$ vicinity on the temperature increase. The latter is related with the decrease of pinning force at $T>T_c$ and with a gradual tearing off vortices from pinning-centers which is manifested in the increase of dissipation. This process, while approaching to $T_c$, is changed on the melting process of Abrikosov vortex lattice near $T_c$ and by the vortex structure disappearance at $T>T_c$ what resulted in a sharp increase of oscillation period t and, correspondingly, to the same sharp decrease of damping of oscillations in a close vicinity of $T_c$.

As it was above noted in the investigated by us system $Bi_{1,7}Pb_{0,3}Sr_2Ca_2Cu_3O_{10-\delta}$ the critical temperature of superconductive transition is equal to $T_c$=107 K.

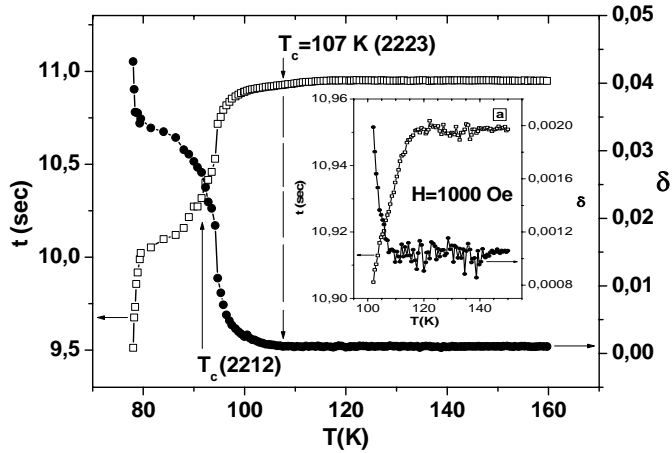

Fig.3. The dependence of period and the logarithmic decrement of dissipation of oscillations of high-temperature superconducting sample Bi-Pb-Sr-Ca-Cu-O system on temperature in a static magnetic field H=1000 Oe.

But results for one of Bi-Pb-Sr-Ca-Cu-O samples, presented in fig.3 show the presence in the studied sample both the more low-temperature phase, then the (2212) phase with $T_c$=95 K, and a more high-temperature, new, unknown nature phase, then superconductive phase (2223) with $T_c$=107 K. It should be paid the particular attention on the dependence character near and below $T$=83 K, what in the revelation of some new low-temperature phase. As for a high-temperature phase of unknown nature, which is higher in temperature, then the main superconductive phase (2223), at first glance it could be related with non-superconducting magnetic phase (because so far it was not observed the presence of Meissner effect due to a low sensitivity of used by us a standard method for the Meissner effect study), but as it will be shown below, the given phase behave in completely other manner then one expects from an usual magnetic phase.

In Fig.3 and particularly in fig.4 in the temperature interval $T$=107-138 K it is clearly seen the ripples both the period and dissipation of oscillations of a sample, and this temperature interval is named by us as the "Chaos" region. A typical dissipation $\delta$ maximum correlating with a sharp change (increase) of the oscillation period $t$, is changed at $T > T_{Ch}^{max}$=138 K, by the exit of period t on the plateau (see in Fig.4) what is characteristic for a materials transforming from the superconducting to the normal state. The comparison of the region ($T$=130-145 K) marked out by frames (a) with dependences in Fig.2 one could see a full analogy. And this is that peculiarity of a new magnetic phase of unknown nature which was discussed above.

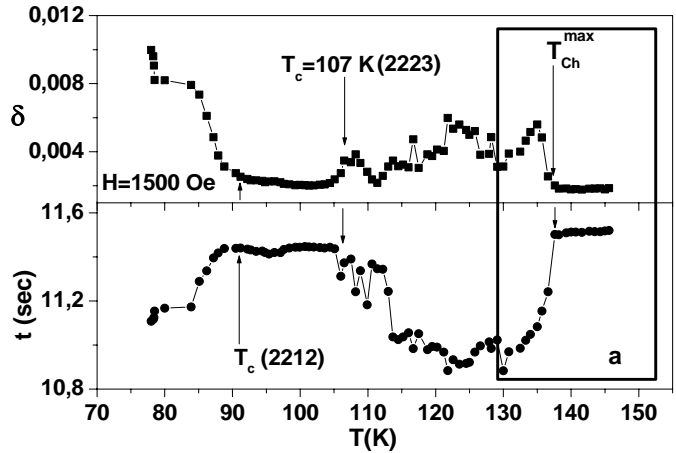

Fig.4. The dependence of period *t* and the logarithmic decrement of dissipation $\delta$ on temperature of strongly anisotropic high-temperature superconductor $Bi_{1.7}Pb_{0.3}Sr_2Ca_2Cu_3O_{10-\delta}$ system in the static magnetic field *H*=1500 Oe.

The results presented in figures 3 and 4 were obtained on different samples of the Bi-Pb-Sr-Ca-Cu-O system.

All these makes it possible to suppose the presence of precursors of superconductive regions in the investigated by us HTSC samples up to the 138 K temperature. This supposition is confirmed by measurements of our sample's resistance temperature dependence at the transition into superconducting state where besides the 2223 phase it is clearly seen the presence of other phases [31].

And, finally, the curve presented in Fig.5 shows that the increase of the outer magnetic field displace in the temperature respect the upper limit $T_{Ch}^{max}$ of the "chaos" region. The ripples of period t and oscillations dissipation $\delta$ at the increase of magnetic field *H* from 1500 Oe up to 20000 Oe are observed up to *T*=150 K, i.e. $T_{Ch}^{max}$ observed at *T*=138 K at *H*=1500 Oe, is displaced up to $T_{Ch}^{max}$=150K.

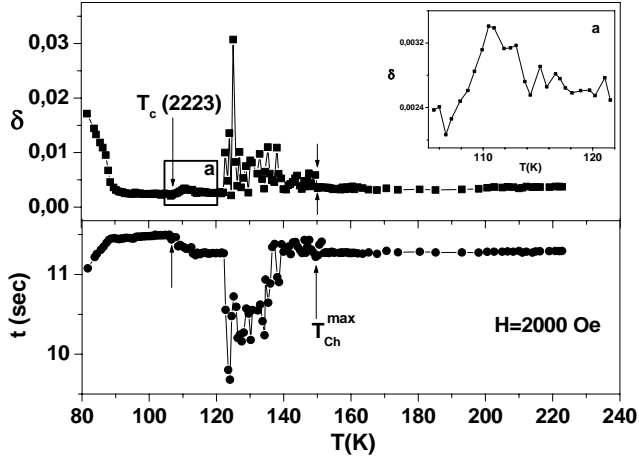

Fig.5. The dependence of period *t* and the logarithmic decrement of dissipation $\delta$ of oscillations for strongly anisotropic high-temperature superconductor of $Bi_{1.7}Pb_{0.3}Sr_2Ca_2Cu_3O_{10-\delta}$ system on temperature in the static magnetic field *H*=2000 Oe.

It should be noted also that for the existence of rudiment of superconducting regions apparently is small because we could manage to record them only using highly sensitive mechanical method of investigations of dissipation processes in high-temperature superconductors.

## Conclusions

In strongly anisotropic high-temperature superconductive samples of the Bi-Pb-Sr-Ca-Cu-O system, synthesized using the solar energy and superfast melt quenching, it was observed the "chaos" region which could probably show to the presence of other high-temperature magnetic or superconductive phases with higher critical temperatures (then the existing in sample of the main HTSC Bi (2223) phase) as separate islands in the normal metal matrix. The determination of these concrete phases and the increase of their percentage content in samples could result in the essential increase of critical temperature $T_c$ of superconducting transition.

**Acknowledgement**

The work is supported by the grants of International Science and Technology Center (ISTC) G-389, G-593 and STCU #4266.